# A High Spectral Resolution Observation of the Soft X-ray Diffuse Background with Thermal Detectors

D. McCammon, R. Almy[1], E. Apodaca, W. Bergmann Tiest[2], W. Cui[3], S. Deiker[4], M. Galeazzi, M. Juda[5], A. Lesser, T. Mihara[6], J. P. Morgenthaler, W. T. Sanders and J. Zhang[7]

Physics Department, University of Wisconsin, Madison, WI 53706, USA; mccammon@wisp.physics.wisc.edu

AND

E. Figueroa-Feliciano, R. L. Kelley, S. H. Moseley, R. F. Mushotzky, F. S. Porter, C. K. Stahle, and A. E. Szymkowiak[8]

NASA Goddard Space Flight Center, Greenbelt, MD 20771, USA

## ABSTRACT

A high spectral resolution observation of the diffuse X-ray background in the 60 - 1000 eV energy range has been made using an array of thirty-six 1 mm$^2$ microcalorimeters flown on a sounding rocket. Detector energy resolution ranged from 5–12 eV FWHM, and a composite spectrum of ~1 steradian of the background centered at $l = 90°$, $b = +60°$ was obtained with a net resolution of ~9 eV. The target area includes bright 1/4 keV regions, but avoids Loop I and the North Polar Spur. Lines of C VI, O VII, and O VIII are clearly detected with intensities of $5.4 \pm 2.3$, $4.8 \pm 0.8$, and $1.6 \pm 0.4$ photons cm$^{-2}$ s$^{-1}$ sr$^{-1}$, respectively. The oxygen lines alone account for a majority of the diffuse background observed in the *ROSAT* R4 band that is not due to resolved extragalactic discrete sources. We also have a positive detection of the Fe-M line complex near 70 eV at an intensity consistent with previous upper limits that indicate substantial gas phase depletion of iron. We include a detailed description of the instrument and its detectors.

*Subject headings:* instrumentation: detectors — instrumentation: spectrographs — intergalactic medium — ISM: general — space vehicles: instruments — X-rays: diffuse background — X-rays: general

## 1. INTRODUCTION

Many years of observations with ever-increasing spatial resolution and statistical detail have shown that the soft X-ray diffuse background is produced by a complex combination of sources, and that observations of this radiation have the potential to convey a great deal of information about a variety of systems. The Galactic interstellar medium is now known to contain substantial quantities of hot X-ray emitting gas, and may in fact be dominated by this component. Most of the luminous mass in clusters of galaxies is also hot gas, and hot clumps and filaments in intergalactic space may be the repository for a majority of present-day baryons (Cen & Ostriker 1999). Distant AGN are another important contributor that dominates at 1-2 keV energies (Mushotzky et al.

current addresses:
[1]Beloit College, Beloit, WI 54555; [2]SRON, 3584 CA Utrecht, The Netherlands;
[3]Physics Dept., Purdue Univ., West Lafayette, IN 47907; [4]NIST, 325 Broadway, Boulder, CO 80305
[5]CfA, 60 Garden St., Cambridge, MA 02138; [6]RIKEN, Wako, Saitama 351-01, Japan;
[7]W.E.S.T., 12 Hawthorne Ave., Adington, MA 02476, [8]Yale Univ., New Haven, CT 06520.

2000). This multiplicity of sources, however, makes it difficult to disentangle and characterize any one of them. Some progress has been made through shadowing experiments and other evidence of interaction with interstellar gas, but even the tremendous boost in the statistical quality and angular resolution provided by *ROSAT* has only added to the richness and complexity of the problem (Burrows & Mendenhall 1991; Snowden, McCammon, & Verter 1993; Snowden et al. 1997).

Another observational approach to the diffuse X-ray background that has yet to be exploited to any significant extent is high resolution spectroscopy. Radiation from gas in the $10^6 - 10^7$ K temperature range where soft X-ray emission is important is entirely dominated by lines of the partially ionized metals, even when their abundance is a small fraction of solar. Observation of large numbers of these lines provides both detailed information about the physical state of the emitting material and enough redundancy to allow superposed sources or nonuniform conditions to be detected and analyzed. Comparison of electron temperature with ionization state can reveal much about the heating history of the gas, and distant extragalactic sources can be separated by redshift.

A few early diffuse background observations with gas scintillation counters and solid state detectors provided tantalizing evidence that emission lines existed in the 500 – 1000 eV range (Inoue et al. 1979; Schnopper et al. 1982; Rocchia et al. 1984), while recent observations with CCDs provide convincing detail (Gendreau et al. 1995; Mendenhall & Burrows 2001). The highest resolution observation up to now was made with the large Bragg crystal Diffuse X-ray Spectrometer (DXS) experiment carried as an attached Shuttle payload (Sanders et al. 2001). It covered the 150 – 300 eV energy range with 5 – 15 eV FWHM resolution, which still leaves most of the lines in this particularly crowded part of the spectrum resolved only as blends. This observation was made in the Galactic plane, where a single nearby component is expected to dominate at these energies, and it showed rather conclusively that the simple collisional equilibrium models traditionally used to fit observations of this thermal emission are not even qualitatively realistic. Solving this problem will require either better astrophysical models and improved atomic physics data, or resolution adequate to resolve lines individually so that analysis can be based on particular features that are well understood, as is commonly done in optical and UV spectroscopy.

The lack of high resolution spectroscopy in the X rays has been largely an instrumental problem. Dispersive instruments can have excellent energy resolution, but have small throughputs and are generally unsuited for extended sources. As an example, the DXS spectrometer had 2500 cm$^2$ total detector area and a 15° × 15° field of view and was carefully optimized for a diffuse source, but detected less than one photon per second. Without the Bragg crystals, the counting rate would have been more than $10^3$ times higher. Energy-resolving ("non-dispersive") detectors, such as solid state diodes and proportional counters, have large acceptance angles and high throughputs, but are fundamentally limited in energy resolution by the statistics of the ionization process to values that are inadequate for many astronomical spectroscopy problems. In this paper we describe a new kind of detector, the single-photon microcalorimeter, that combines spectral resolution comparable to dispersive instruments with the high throughput and efficiency of a non-dispersive detector, and present some of its first results. These devices are now scheduled to be flown on the Astro-E2 observatory (Kelley et al. 1999), and are baselined for future X-ray missions such as Constellation-X (see http://htxs.gsfc.nasa.gov/) and XEUS (see http://astro.esa.int/SA-general/Projects/XEUS/).

A general description and characterization of microcalorimeter detectors is given in §2. Section 3 describes the sounding rocket instrument, §4 the flight observations, and §5 outlines the data reduction and problems encountered. The science results are discussed in §6. We point out here that the worst of the difficulties encountered in the instrument design and data reduction are unique to the particularly hostile environment of a sounding rocket flight, and should be much less serious for a satellite instrument where the disturbances are considerably smaller and available recovery times longer. On the other hand, future generations of detectors will have higher resolution and be proportionately more sensitive to all of these problems, so we can expect them to surface again on some future mission and it will be well to have faced them and learned the appropriate lessons.

## 2. MICROCALORIMETER DETECTORS

Calorimeters are conceptually simple devices, consisting of only three components: an absorber large enough to intercept and contain the energy to be measured, a thermometer to measure the resulting temperature rise of the absorber, and a weak thermal link to a heat sink to cool the absorber back to its starting point. They have been used since the early days of nuclear physics to measure the integrated energy of various radioactivities, and by the mid 1930s Kurti and Simon had demonstrated the sensitivity advantages of cryogenic operation, running a small calorimeter at 0.05 K to measure a weak source (Simon 1935). Modern technology has made both detector construction and cryogenics more convenient, but there have been no really essential developments, and Kurti's device must have been very nearly capable of detecting individual particle or gamma ray events. However, the earliest published references that we have found to this possibility are a note on the detection of transitions produced by alpha particles in an NbN superconducting strip (Andrews, Fowler, & Williams 1949), and the 1974 account by Tapio Niinikoski of spurious pulses on a carbon resistance thermometer readout, which he identified with local heating due to the passage of individual cosmic rays (Niinikoski 1975). Shortly thereafter, he proposed a calorimetric neutrino pulse detection experiment at CERN (Niinikoski 1974). This was never implemented, so the first experimental efforts did not begin until 1982, when Niinikoski helped Ettore Fiorini of Milano design large calorimetric detectors to expand the range of materials that could be investigated for neutrinoless double beta decay (Fiorini & Niinikoski 1984). Coincidentally, this was about the same time that Harvey Moseley at Goddard Space Flight Center independently suggested to the X-ray group there that thermal detection should offer much better energy resolution than any semiconductor alternative for astronomical X-ray spectroscopy. A Goddard/Wisconsin collaboration soon demonstrated the possibilities of this technique (Moseley, Mather, & McCammon 1984; McCammon et al. 1984) and has since developed both the sounding rocket experiment discussed here and the X-ray spectrometer for Astro-E (Kelley et al. 2000).

Solid state detectors, proportional counters, and CCDs all work by detecting the ionization produced following absorption of a photon. Only a fraction (~1/3) of the deposited energy goes into ionization, and the resolution of these devices is fundamentally limited by statistical fluctuations in this branching ratio. These can be expressed as $\sqrt{fWE}$, where $E$ is the deposited energy, $E/W$ is the average number of ionizations produced by $E$, and the Fano factor $f$ accounts for the suppression of Poisson fluctuations due to the finite ionization energy (Fano 1947). If this ionization energy were negligible, the ionizing events would be statistically independent, and the fluctuations would be determined by Poisson statistics with $f = 1$. If the ionization efficiency were 100% (W = ionization energy), there would be no fluctuations and $f = 0$. For silicon, $W = 3.65$ eV per ionization (Janesick et al. 1988), the ionization efficiency is about 33%, and $f = 0.115$ (Mendenhall et al. 1996). This gives a theoretical limiting resolution of 118 eV FWHM at $E = 6000$ eV.

Thermal calorimetry on the other hand is in principle a quasi-equilibrium measurement with no branching and thus no fundamental limitation on resolution. The random exchange of energy carriers between the absorber and heat sink through the thermal link produces fluctuations in the energy content of the absorber whose rms magnitude is $\sqrt{kT^2C}$, independent of the characteristics of the link. These do not impose any direct limitation on the accuracy of determining the deposited energy, however, since they occur on a characteristic time scale $\tau = C/G$, where $C$ is the heat capacity of the absorber and $G$ the thermal conductivity of the link, and it is only necessary to measure the change in temperature due to the absorbed photon quickly relative to this to achieve arbitrary accuracy. A limitation comes in when the intrinsic noise associated with a real thermometer is considered. The thermometer reading must be averaged for a finite time to limit this noise, and some energy fluctuation over the link is then included. Better thermometers have less internal noise for a given sensitivity to temperature, and so can measure faster and therefore more accurately.

At this point it might seem that it is only necessary to decrease $G$ and slow the fluctuations down to improve the resolution. However, most thermal detectors currently use thermistors (semiconductor or superconducting transition edge), which require bias power to read out the resistance changes. These have an output signal proportional to readout power, while the fundamental Johnson noise of the resistance is constant, so signal-to-noise ratio improves with increasing bias. But the readout power raises the absorber temperature and rapidly increases the magnitude of the energy fluctuations, so there is an optimum bias which depends weakly on thermometer and absorber parameters but is always proportional to $G$ and results in a temperature rise of roughly 15% above the heat sink. Decreasing $G$ requires a proportional reduction in bias power to keep the same temperature rise, and the resulting reduction in responsivity for a fixed Johnson noise requires a longer averaging time, exactly canceling the gain from the longer fluctuation time scale. For an ideal resistive thermometer there is therefore no dependence of resolution on speed. This is not generally true for real devices, where thermometers have limited power density capabilities and deposited energy takes a finite time to thermalize.

Problems of ensuring adequate thermalization require care in choice of materials. There must be no significant escape mechanisms or metastable states that can trap energy for significant lengths of time. Thermalization and equilibration with the thermometer should ideally be complete within the required thermometer averaging time. This requires that the thermalization time be less than $\tau/\alpha$, using a resistive thermometer with figure of merit $\alpha \equiv d\log(R)/d\log(T)$. For detectors of high energy photons, the problem of providing an absorber with a high stopping efficiency over a large collecting area without too much heat capacity further complicates the choice of materials.

There are a number of publications with extensive general descriptions of microcalorimeter detectors and their applications (Stahle, McCammon, & Irwin 1999; Stahle 1996; McCammon et al. 1993; Wollman et al. 1997; Cabrera et al. 1998), and many on the details of various aspects of their design and operation (Kelley et al. 1993; McCammon et al. 1999; Stahle et al. 1996; Zhang et al. 1993; Zhang et al. 1998; Stahle et al. 1999). Some of the best resolution results to date are ~ 4.5 eV FWHM for 6 keV X-rays (Alessandrello et al. 1999; Irwin et al. 2000; Bergmann Tiest et al. 2002), which is a factor of 26 better than the theoretical resolution of a silicon solid state detector or CCD, and 0.15 eV FWHM for visible photons (Romani et al. 1999).

3. THE X-RAY QUANTUM CALORIMETER SOUNDING ROCKET INSTRUMENT

In this section we describe the integration of microcalorimeter technology into a suborbital sounding rocket instrument. We have gone to considerable effort to overcome the challenges of adapting this inherently sensitive and quiescent technology to the inherently rough and time-constrained sounding rocket environment because we felt it essential a) to gain experience with operating these detectors in space, b) to do important diffuse-source science that is not possible with any currently scheduled mission, and c) to provide a flight test-bed for more advanced cryogenic detector technology. The instrument is composed of thirty-six 1 mm$^2$ detectors operated at 0.06 K using pumped liquid helium and a paramagnetic demagnetization refrigerator. The detectors with their thin IR/vis/UV blocking filters have a ~1 sr field of view and are optimized for observing the soft X-ray background in the region below 1 keV, although the payload can also be configured with the detector array in the focal plane of an imaging conical-foil mirror with about 250 cm$^2$ effective area (Serlemitsos & Soong 1996). The following sections describe each of the instrument components in some detail, and include performance data from the flight. Much of this performance data is relevant to the feasibility of future missions with more sensitive detector systems.

3.1. *The Cryostat*

Applying these new detectors to X-ray astronomy requires of course that they be operated in space. Providing the requisite low operating temperatures in a sounding rocket environment presents a particularly difficult technical challenge due to the high vibration levels and short time scales. The payload is subject to about 17 g's r.m.s. vibration during the 40 s of powered flight, and reaches observing altitude only 80 s after motor burnout. Cooling is provided by a magnetic refrigerator that requires at least half an hour for the cooldown portion of its cycle, which therefore must be completed before launch.  The cryostat design is then driven by the need for sufficient isolation from launch vibration that mechanical heating does not exceed the limited total cooling capacity of the refrigerant, and for sufficiently short thermal time constants that very precise temperature regulation at 60 mK can be re-established within a few tens of seconds after burnout.

      A cross-section of the cryostat is shown in Figure 1.  It consists of a 4 L annular liquid helium container suspended from the vacuum jacket on two reentrant fiberglass-epoxy cylinders, with a single straight fill/pumping line that contains two vapor-cooled heat sinks connected to radiation shields (Steffensrud et al. 1991). The spaces between the shield and vacuum jacket and between the two shields are filled with aluminized polyester multilayer insulation.  The outer shield runs at 130 K, and the inner at ~25 K, with the pumped He bath at 1.6 K.  The helium hold time is about 28 hours.  This would be improved by a factor of about three if the inner shield continued up the inside of the inner epoxy shell and across the front of the dewar between the 130 K and 1.6 K shields, but this was precluded by the requirements for a 1 steradian field of view and a reasonably small diameter for the outer filter, which necessitated very close spacings at the front end.

      Detector cooling at 60 mK is provided by an adiabatic demagnetization refrigerator (ADR) with a 50 g iron-alum salt pill suspended on aramid (Kevlar) fibers in the bore of a 4 Tesla superconducting solenoid.  An ADR is an almost ideal cooler for space applications, since it is all solid, requires no moving parts, and can be made to operate at near Carnot efficiency (Pobell 1996).  The complete ADR assembly with magnet weighs less than 2 kg and is mounted in the central hole of the helium tank.  The refrigerator will maintain a 60 mK cold plate temperature for about 12 hours on a single cycle.

      The thermal time constant between the cold plate and salt pill is about one second, allowing fast closed-loop control of the temperature.  Two identical nuclear transmutation doped germanium resistance thermometers (Haller et al. 1984) are mounted on the back of the cold plate.  One is used for control, the other as a monitor.  Temperature control circuitry is greatly simplified by controlling the magnet terminal voltage — equivalent to $dI/dt$ — rather than the magnet current. The temperature error signal from the control thermometer is fed through a simple analog PID controller that uses three op-amps to generate terms proportional to the error and its time derivative and integral, and applies this voltage to the magnet terminals.  Cold plate temperature can be controlled to about 160 nK r.m.s. at 60 mK, and to 35 nK r.m.s. at 50 mK.  This is much better than the ~20 µK required for our current detectors, but will be useful for next-generation devices with more sensitive superconducting thermometers.

      The greatest challenge in the cryostat design was isolation from the launch vibration.  For the Black Brant V sounding rocket, the flight-level specification is about 0.1 $g^2$ $Hz^{-1}$ from 5 Hz to 2000 Hz, or about 17 g's r.m.s, which lasts for 30 s.  Modeling a suspended mass as a damped harmonic oscillator, it can be shown that the heat input from a flat acceleration spectrum like this is independent of both the resonant frequency and the Q, and depends only on the mass. The relevant constant is 13 W $kg^{-1}$ $g^{-2}$ Hz, which for our 150 gram cold stage would predict ~200 mW vibrational heat input with no isolation, compared to <1 µW for the normal parasitic load.  To avoid exhausting the cooling capacity of the salt pill, we therefore require at least a factor of 200 isolation.

      This isolation is provided by staggering the resonant frequencies of the various structural elements, as shown in Figure 2. The entire cryostat, which weighs 26 kg including most of the analog electronics, is suspended from the rocket skin on commercial elastomer vibration dampers. These give a resonant frequency of 40 Hz and a Q of about 3.  The resonant frequency of the helium can on its reentrant cylinders is 100 Hz. The cold stage is suspended kinematically on six Kevlar fibers, which have a very high ratio of stiffness to thermal conductivity below 20 K.  The lowest frequency modes of this suspension are all ~400 Hz, and the predicted isolation factor is >1000. Vibration tests of the completed system verified the model, with an actual isolation about a

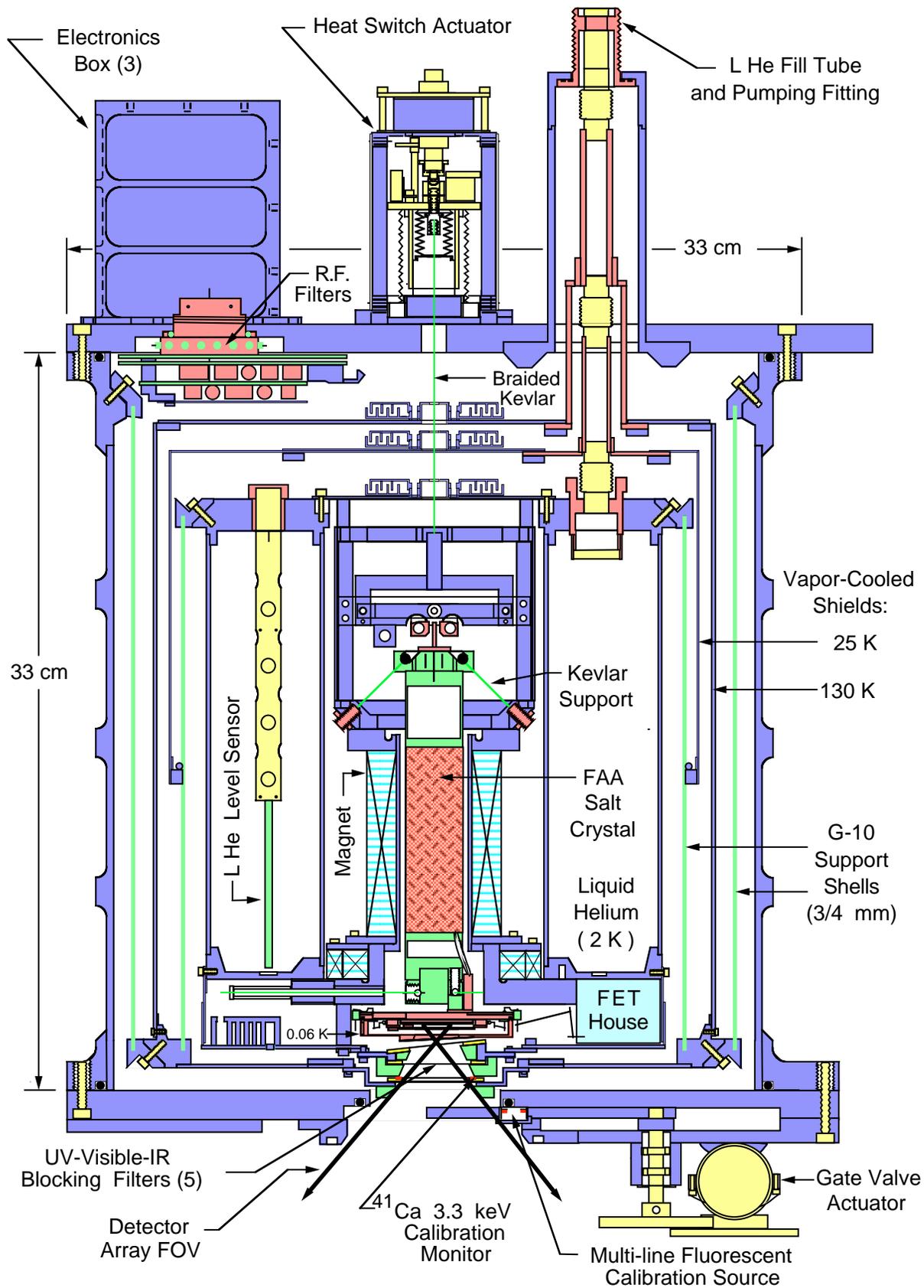

Fig. 1.—Cross section of sounding rocket cryostat. Total weight is 27 kg with electronics and vibration mounts.

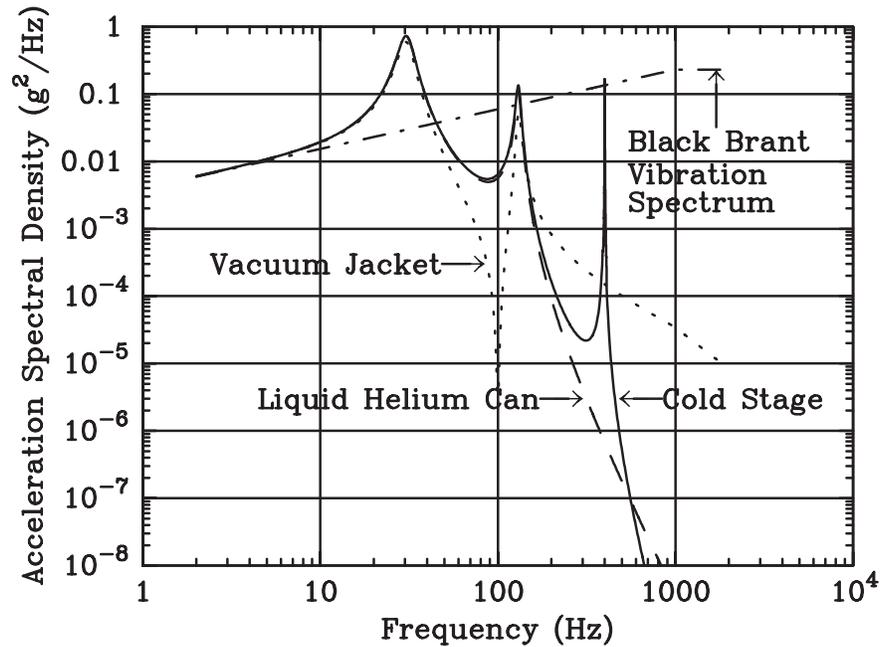

Fig. 2.—Spectral density of vibration input from rocket motor and response of various components of the refrigerator. The response is higher than the input level at the resonant frequency of the stage, but drops rapidly above this. Staggered resonant frequencies provide > 1000 × isolation from launch vibration at the cold stage resonance to limit heat input at 60 mK.

factor of two better than predicted (Cui et al. 1994). The vibration environment on a satellite observatory will be much less severe, but cannot be ignored. We have described the problem and our approach to solving it in some detail because similar measures will be required to prevent low-level vibrations, such as produced by reaction wheels and gyroscopes, from degrading the resolution of high performance detectors.

### 3.2. *UV/vis/IR Blocking Filters*

Long wavelength photons that are individually below the detection threshold can have two undesirable effects. They heat the detectors, reducing their sensitivity, and their shot noise adds to the detector noise. The second problem is usually more serious, with a crossover at about 1 mm wavelength, where shorter wavelengths produce shot noise effects before they cause appreciable heating, and longer wavelengths can heat the detector significantly without adding to the noise. Thermal infrared radiation from a 300 K background must be attenuated by a factor of $\sim 10^9$ for wavelengths between 2 μm and 2 cm or it will degrade the detector performance for our devices.

We would like to observe soft X rays down to the 100 eV range with good efficiency, but no material exists with a sufficiently high ratio of absorption coefficients between 0.5 eV and 100 eV. We instead make use of the very large real index of refraction of aluminum in the infrared to reflect away the long wavelength radiation. It is advantageous to divide the aluminum into several thin layers, since infrared has large reflection losses at each surface, while X rays are relatively unaffected by surfaces. We employ five 150 Å aluminum filters in series, each supported on a 1380 Å parylene substrate and tilted ~ 3° with respect to the others to minimize multiple reflections. This number is approximately optimum, since both the substrates and the oxide layer that forms on the aluminum surfaces contribute additional X-ray absorption. Infrared transmission measurements of individual filters are in good agreement with calculations based on bulk optical constants for aluminum. It is impractical to measure the very small effective transmission of the stack, so a Monte Carlo ray tracing program that includes the effects of multiple reflections and thermal emission from the filters is used to verify the net attenuation. The parylene serves also to attenuate the very bright 41 eV He II Ly-α geocoronal line, which is not strongly absorbed by aluminum.

### 3.3. *Calorimeter Array*

The detector assembly is a 2 × 18 pixel array micromachined from a single piece of silicon. Each of the pixels is supported on three thin legs for thermal isolation (Stahle et al. 1996, 1999). Doped silicon thermistors are incorporated into the pixel structure by ion implantation and electrical connections are provided by degenerately doped ion-implanted tracks that run down the centers of the legs to bonding pads at the periphery of the chip. The construction and characterization of this type of thermometer has been discussed by Zhang et al. (1993, 1998). The 1 mm$^2$ HgTe absorbers were attached using a small silicon spacer and Stycast 2850FT epoxy as shown in Figure 3. This absorber provides better than 99% stopping efficiency for X rays below 1 keV. Detector dimensions and heat capacity budget are given in Table 1.

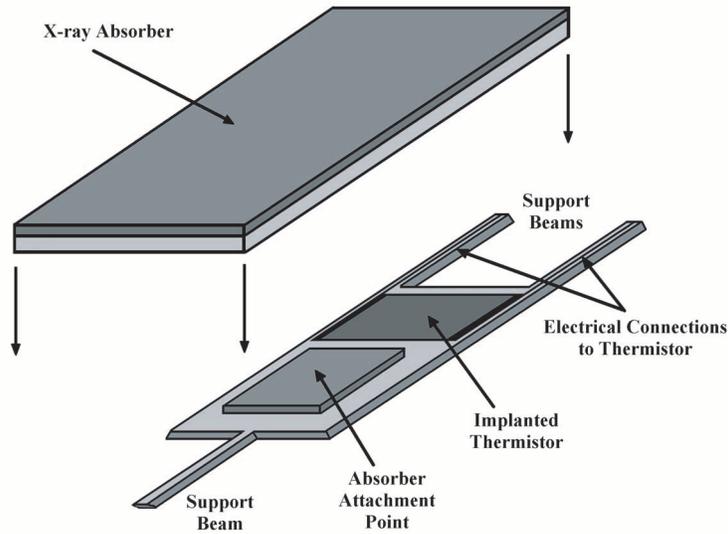

FIG. 3.—Detector pixel assembly. The HgTe X-ray absorber on a 14 μm silicon substrate is attached to the small silicon spacer and the spacer to the pixel body with epoxy. Dimensions are given in Table 1.

A 100 μm thick tungsten mask shields the pixel legs and frame of the array from X-ray hits. The absorbers overhang the gaps between the detectors and shield the segment of the third leg that runs between the pixels. Figure 4 is a photograph of the detector array in the inner housing before the tungsten shield is installed. The detector thermal and operating point parameters are summarized in Table 2. The R(T) function given there is an empirical fit to the systematic deviations from simple variable range hopping behavior discussed in Zhang et al. (1993).

Because of the small G, the detectors are sensitive to heating by sources of stray power as small as $10^{-14}$ W. We initially had problems with heating that were apparently due to millimeter-wavelength thermal radiation conducted into the detector enclosure through the electrical feedthroughs, re-radiated inside the enclosure, and absorbed by the HgTe on the detectors. Improvements in the feedthrough design helped, but we were forced to use two nested detector boxes to eliminate the problem (Bergmann Tiest 1999). The electrical feedthroughs are 4.8 kΩ RN50c-style nichrome film resistors that are glued into holes in the 1.5 mm copper walls of the detector enclosures with conductive silver epoxy. These are very simple and work as well as any of several feedthrough designs we tested.

TABLE 1
Pixel Physical Characteristics

| Component | Dimensions | Heat Capacity Budget (J/K at 70 mK) |
|---|---|---|
| HgTe Absorber | 0.5 mm × 2.0 mm × 0.96 μm | $1.20 \times 10^{-14}$ |
| Si Absorber Substrate | 0.5 mm × 2.0 mm × 14 μm | $0.43 \times 10^{-14}$ |
| - Surface excess heat capacity | | $1.75 \times 10^{-14}$ |
| Si Detector Body & Isolation Legs | | $0.06 \times 10^{-14}$ |
| - Detector Body | 0.25 mm × 1.0 mm × 7 μm | |
| - Absorber Mounting Spacer | 245 μm × 245 μm × 12 μm | |
| - Short Thermal Isolation Legs (2) | 900 μm × 12 μm × 7 μm × 1/3 | |
| - Long Thermal Isolation Leg | 2900 μm × 20 μm × 7 μm × 1/3 | |
| - Surface excess heat capacity | | $0.50 \times 10^{-14}$ |
| Ion-implanted Si Thermistor ($6 \times 10^{18}$ cm$^{-3}$ P, $3 \times 10^{18}$ cm$^{-3}$ B) | 200 μm × 400 μm × 0.23 μm | $1.50 \times 10^{-14}$ |
| Stycast™ Epoxy (2 spots) | (25 μm)³ | $0.56 \times 10^{-14}$ |
| Contact & Trace Degenerate Implants | | $0.55 \times 10^{-14}$ |
| Total Heat Capacity Budget | | $6.55 \times 10^{-14}$ |
| Measured Heat Capacity | | $6.00 \times 10^{-14}$ |

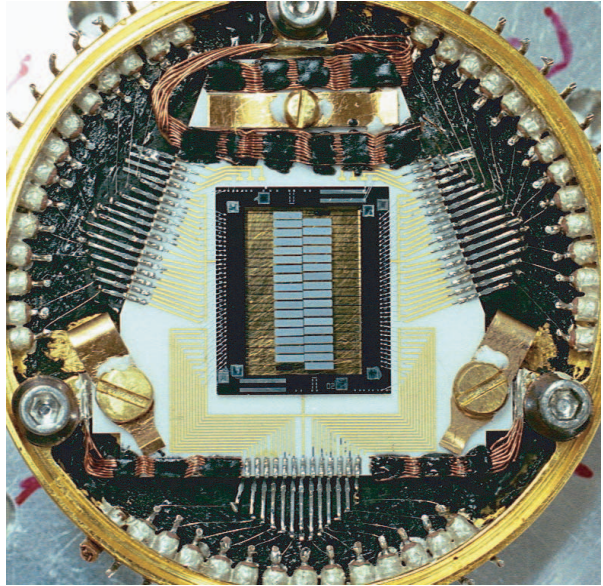

FIG. 4.—36-pixel array in inner detector box. The white alumina board is kinematically mounted on three brass spheres to avoid thermal stresses. A tungsten mask was added later to shield the pixel legs.

TABLE 2
Pixel Thermal and Operating Point Parameters

| Parameter | Value |
|---|---|
| $G_{isolation\ legs}$ | $1.2 \times 10^{-11}$ $(T/0.1\ K)^{2.1}$ W K$^{-1}$ |
| $G_{electron-phonon}$ | $2.2 \times 10^{-10}$ $(T/0.1K)^{4.4}$ W K$^{-1}$ |
| $T_0$ ($\pm$ variation over array)$^a$ | $5.36 \pm 0.04$ K |
| $R_0$ ($\pm$ variation over array)$^a$ | $880 \pm 14$ $\Omega$ |
| $T_{cold\ plate}$ | 59 mK |
| $T_{detector\ pixel}$ | 70 mK |
| $V_{bias}$ | 8 mV |
| $R_{load\ resistor}$ | 88.5 M$\Omega$ |
| $R_{detector\ pixel}$ (at bias point) | 7.8 M$\Omega$ |
| $P_{bias}$ | $5 \times 10^{-14}$ W |
| C | $6.0 \times 10^{-14}$ J K$^{-1}$ |
| G | $5.6 \times 10^{-12}$ W K$^{-1}$ |
| $\tau_{thermal}$ | 10.7 ms |
| $\tau_{effective}$ | 7.2 ms |
| $\alpha$ (= d(log R)/d(log T) ) | $-4.4$ |
| NEP(0) | $1.3 \times 10^{-18}$ W Hz$^{-1/2}$ |
| Field of View | 61° (0.81 steradians) |

$^a$ $R(T) = R_0 e^{\sqrt{T_0/T}} + R_0' e^{\sqrt{T_0'/T}}$, where $R_0' = R_0 \exp(2.522 T_0^{1/2} - 8.733)$, and $T_0' = 2.715 T_0 + 1.233$

### 3.4. Calibration Sources

A ring-shaped 2 μCi $^{41}$Ca source located around the inner edge of the outermost blocking filter mount provides a continuous calibration source to track gain changes during the flight. The potassium $K_\alpha$ and $K_\beta$ lines from this electron-capture source are at 3312 and 3590 eV and so do not seriously contaminate the 70 – 1000 eV range being observed, despite the relatively high count rate of 0.5 s$^{-1}$ per pixel. A multi-target fluorescent source excited by a 1 mCi $^{244}$Cm alpha-emitter is mounted in the center of the gate valve slide and provides lines at 151, 277, 525, 677, 1487, 1740, 2041 and 2124 eV to establish the gain linearity and monitor filter transmission at the beginning and end of the flight. When the gate valve is opened, the source is retracted behind a lead shield, and so does not contaminate the science data.

### 3.5. Electronics

The detectors are approximately current biased through 90 M$\Omega$ load resistors located in the outer 60 mK detector box. The voltage signal across each detector is buffered by a cold junction field effect transistor (JFET) source follower located nearby, then amplified by room-temperature stages attached to the cryostat. The 10-volt full scale pulses are then carried to a separate electronics section, where they are digitized and stored until they can be sent back on the rocket telemetry.

The load resistors are composed of three 30 M$\Omega$ nichrome thin film chip resistors manufactured by Mini-Micro Systems, Inc. We have tested these and find that when operated at 50 mK with as much as 500 pA current, excess noise corner frequencies are below 1 Hz. The first stage amplifiers are silicon JFET source followers operated at 121 K. The JFETs are selected Interfet NJ14ALs with voltage noise levels below 3 nV Hz$^{-1/2}$ and corner frequencies below 3 Hz when operated at $I_d$ = 140 μA, $V_{dg}$ = 1.5 V, and T = 121 K. Current noise was negligible ($<5\times10^{-17}$ A Hz$^{-1/2}$ on tested devices) as long as $V_{dg}$ is kept below 2 V. The JFET chips are mounted in groups of twelve in flatpacks equipped with a chip resistor heater and silicon diode thermometer to regulate the JFET operating temperature. The flatpack is suspended on tensioned Kevlar inside a light-tight aluminum box at 1.6 K with feedthroughs for the twelve gate leads. Three of these "houses" are used for the 36 detectors.

The JFET source followers provide no voltage gain, but drop the impedance to a level where microphonics on long cables is not a problem. They are followed by low noise preamps that use pairs of Toshiba 2SK147 JFETs on the inputs. These provide 1.2 nV Hz$^{-1/2}$ down to 3 Hz or less without selection, and therefore do not contribute significantly to the noise of the cold JFETs, which in turn is negligible compared to the detector noise for thermistor resistances >10 MΩ (a level not quite achieved on this flight due to the decision to operate at 60 mK rather than 50 mK). The preamps operate at room temperature, but are mounted on the inside of the vacuum jacket cover to take advantage of the shielding provided by the vacuum jacket. Every electrical lead going into the dewar includes a monolithic pi-section RF filter. This is essential to avoid RF heating of the detectors and thermometers in most environments.

Three shielded boxes bolted to the dewar cover contain amplifiers with an additional gain of 61, resulting in an X-ray pulse amplitude of about 1 V keV$^{-1}$. These boxes also contain the low-noise power supplies for the detector bias and preamps, and the various housekeeping monitors. All outputs and inputs again go through pi-section filters before going to the cables that connect them to the external digital electronics. Figure 5 shows a block diagram of the analog electronics for one pixel.

In a separate section of the rocket payload, the analog pulses from each pixel are split into two paths. One path is analog filtered to approximately optimize the signal to noise ratio, and the peaks of the resulting pulses are digitized to 12 bits if the signals exceed a first level discriminator at 20 eV. The digitized record can be overwritten by a subsequent pulse unless a second level discriminator is also triggered. This second level alternates between 25 eV and 135 eV at 10 s intervals as insurance against losing all of the flight data to deadtime from an increased noise level. In the other path, the signal from each pixel is digitized continuously at 48 μs intervals with a 12-bit sigma-delta converter and stored in a 1024-word buffer. A 50 ms string of samples containing the pulse and a portion of the pre-trigger baseline that can be digitally filtered on the ground is then telemetered.  Both data paths are triggered at random intervals to obtain unbiased samples of the baseline for analysis of noise, pileup, and deadtime.

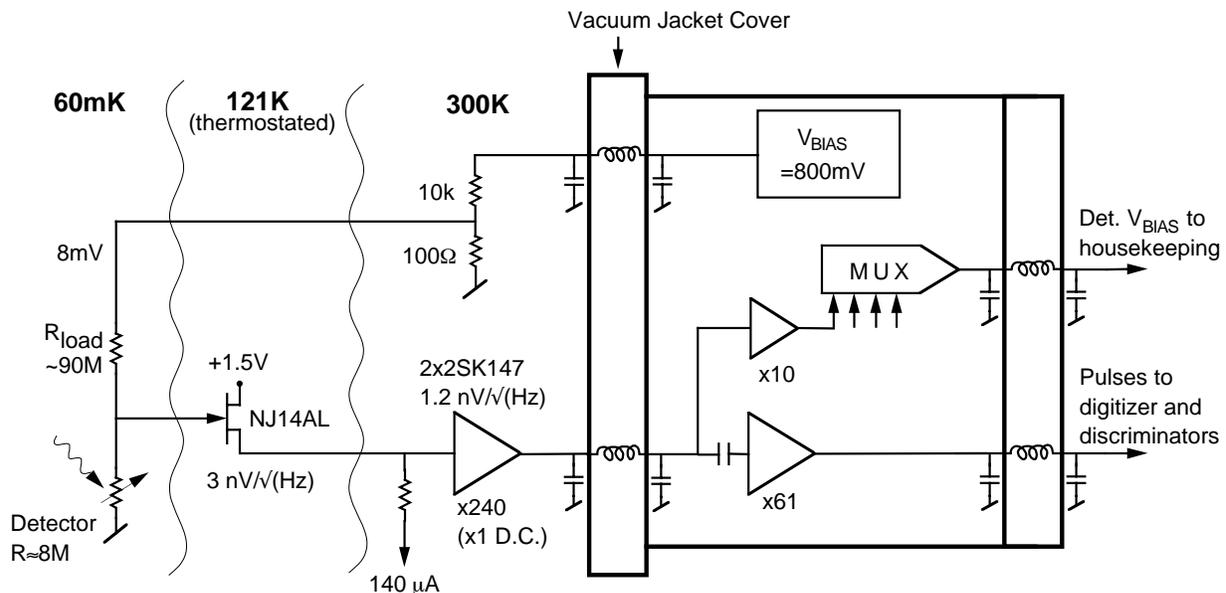

FIG. 5.—Analog electronics chain for a single pixel. These circuits are inside the cryostat or mounted on the top cover.

The data, housekeeping, and other engineering information are sent back to the ground on two 800 kb s$^{-1}$ PCM telemetry channels.  The signal capacity is 32 counts s$^{-1}$ pixel$^{-1}$ for the analog-filtered data and 2 counts s$^{-1}$ pixel$^{-1}$ for the digital data.  To keep the $^{41}$Ca calibration source events from using up a large fraction of the digital telemetry capacity, only every fourth event above 1.6 keV is saved on this path, while all are sent on the analog channel.  The net expected rates are then about 0.5 counts s$^{-1}$ pixel$^{-1}$.

## 4. SOUNDING ROCKET FLIGHT

For this flight, no optics are required.  The outermost blocking filter mount, located 28 mm in front of the array, functions as a field stop and defines a ~1 steradian field of view. Two of the 36 electronics channels were connected to thermometers embedded in the frame of the array to monitor transients in the chip temperature.  Data from the remaining 34 pixels were combined to provide a total collecting area of 0.34 cm$^2$ while keeping the individual detectors small enough to achieve good energy resolution.  The field of view was centered at $l = 90°$, $b = +60°$, a region chosen to be typical of high-latitudes for the 500–1000 eV region by avoiding Loop I, the North Polar Spur, and other features thought to be due to supernova remnants and superbubbles.  About 38% of this general high-latitude background is known to be produced by distant AGN (Hasinger et al. 1993), but the source and emission mechanism of the remainder is unknown.  In the 100–300 eV range, this part of the sky includes some of the brightest high-latitude enhancements.  These apparently are partly due to extensions of the local hot bubble in these directions, but may also be produced by patches of hot gas in the halo (Kuntz & Snowden 2000).  The field of view and path of a 360° scan through the earth to evaluate background are shown in Figure 6.

Flight 27.041UG was launched 1999 March 28 at 09:00 UT from White Sands, New Mexico.  To maintain the helium bath temperature at 1.6 K prior to launch, the vapor over the liquid helium was pumped through a flexible bellows and a port in the rocket skin equipped with a remotely operated butterfly valve.  The ADR cooldown cycle was completed at T–2 hours, and temperature regulation established at 60 mK.  About 45 minutes of calibration source data were recorded prior to the launch.  The behavior of the coldplate temperature is shown in Figure 7.  The operating temperature was recovered to within 10 µK about 10 s after regulation was re-enabled at +60 s, and subsequent variations were ±210 nK r.m.s. during most of the data taking interval. The largest disturbances before re-entry are due to the operation of the gate valve, whose motor was mounted on the vacuum jacket, bypassing a major stage of the mechanical isolation system.

Near peak altitude, the center of the field of view was scanned a full 360° at about 6° s$^{-1}$ along a great circle that passed through the northwestern horizon and the nadir, returning to its original position as shown in Figure 6.  The large-amplitude low-frequency buffeting during reentry exceeded 20 g and produced too much mechanical heating to maintain regulation, but the temperature was recovered again as soon as it stopped, and several minutes of good calibration data were obtained while the payload was on the parachute.

## 5. DATA ANALYSIS

The data acquired from the instrument include the fully sampled digitized waveforms for each X-ray event plus "quick-look" X-ray pulse heights from the on-board analog pulse shaping, and large quantities of housekeeping information.  The post-flight data analysis consists of categorizing events as good X-rays, calibration, cross talk, pileup, or cosmic ray frame-hits.  The X-ray waveforms from each pixel are then digitally filtered, and a nonlinear energy scale and gain-drift corrections are applied before combining the events into a single spectrum.

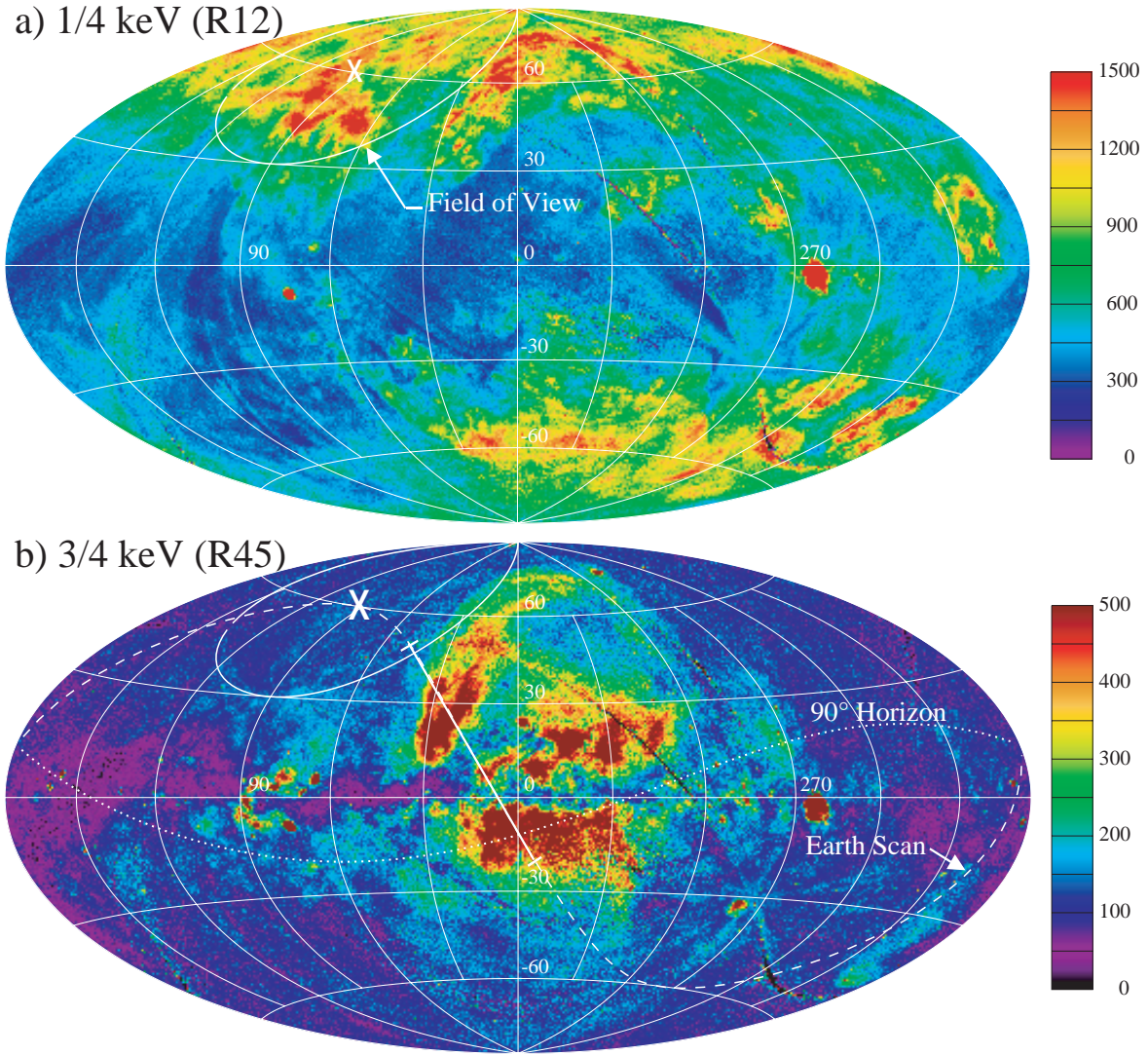

Fig. 6.—Target area projected on *ROSAT* maps of the (a) 1/4 keV and (b) 3/4 keV diffuse background. The ~1 sr field is centered at $l = 90°$, $b = +60°$. A 360° great-circle scan through the nadir as shown in (b) was performed in the middle of the flight at 6° s$^{-1}$ to evaluate the non-X-ray background. *ROSAT* intensity units are $10^{-6}$ cts s$^{-1}$ arcmin$^{-2}$.

## 5.1 Categorization of events

The primary data used in the analysis are the 1024-sample strings containing the unshaped pulses. Typical unfiltered data strings produced by some of the calibration source events are shown in Figure 8. These include about 10 ms of pre-pulse baseline. The level, slope, and noise of this pretrigger part are evaluated for each pulse and later used to categorize the event for pileup on the tail of a preceding pulse or interference that might affect the energy determination. Risetime, falltime, and pulse location relative to the nominal trigger position are also determined and used to screen out extraneous events, which include cross talk, X rays that penetrate the absorber and are stopped in the silicon pixel structure or legs, and cosmic rays. X rays that penetrate the HgTe absorber and are stopped in the silicon absorber substrate lose 3-20% of their energy to electron and hole trapping on impurity states in the bandgap, but the pulse shapes are indistinguishable from HgTe events. We have no way to eliminate these in processing, but the downward shift is limited and the HgTe remains optically thick far enough above 1 keV that there are no substrate
13

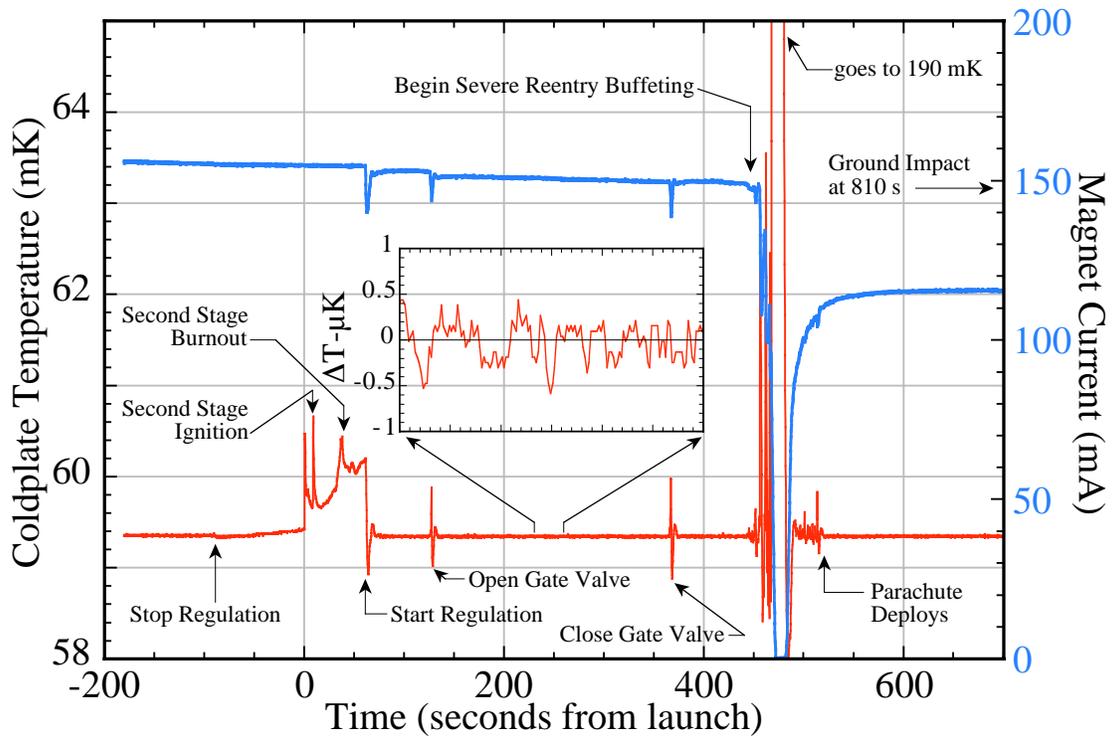

Fig. 7.—In-flight performance of the temperature control system, showing the coldplate temperature and magnet current. Temperature fluctuations during data taking are about 210 nK r.m.s. The gate valve motor is located on the vacuum jacket, and caused the most serious thermal disturbance up to reentry. Accelerations during reentry exceeded 20 g with tumbling at ~1 Hz, introducing heat to the cold stage faster than it could be removed. Temperature regulation is recovered once tumbling stops, allowing calibration data to be obtained.

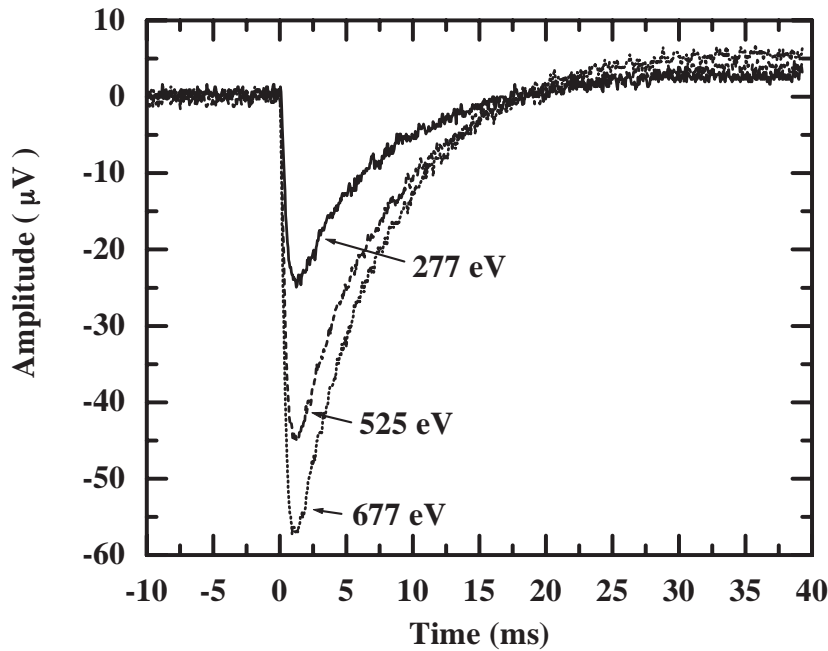

Fig. 8.—Unfiltered X-ray pulses from the gate-valve calibration source.

events below this energy. Overlapping events that peak somewhere within the sample interval are detected by differentiating the data and subtracting the scaled derivative of a clean "template" pulse. Residuals that exceed the noise threshold are flagged as "double hits" and treated as pile-up events.

### 5.2 *Digital Filtering*

For a linear system with stationary noise, it can be shown that the best resolution is obtained with a filter that consists of weighting factors $k_i = \hat{s}_i / n_i^2$, where $\hat{s}_i$ is the complex conjugate of the signal at frequency $f_i$ and $n_i^2$ is the mean square noise at $f_i$ (Boyce et al. 1999). The phase of the weighting factor is reversed from that of the signal so that the filtered signal components will all be real and add in phase, maximizing the amplitude. For white noise, this filter is equivalent to convolving the pulse with a time-reversed template of itself, and the improvement it gives over a causal filter with the same frequency response is 36% for an ideal exponential pulse.

This filter is adaptive in the sense that it uses the average noise observed over some part or all of the data taking period. The digital data storage for each pixel is triggered by an independent free-running oscillator about once every 5 s to obtain a random sample of the system noise. These baseline samples are examined for pileup and accidental pulses, the clean ones Fourier transformed, and the resulting power spectra averaged for the prelaunch part of the data file. We also obtain a template signal pulse by selecting a number of good 3313 eV pulses from the $^{41}$Ca source, averaging these, and Fourier transforming the average to get the signal amplitude and phase to use for the filter weighting coefficients. Separate filters are generated for each pixel. Figure 9 shows the noise spectrum and resulting frequency-domain filter used for pixel 16. Pulse-processing methods are discussed further by Szymkowiak et al. (1993) and by Boyce et al. (1999).

### 5.3 *Gain Normalization*

Despite the very close regulation of the coldplate temperature, the temperature and therefore the gain of the detectors is observed to vary during the flight. There are several causes, some of which we are still working to understand more completely. First, opening the gate valve allows radiation from the warm rocket skin that passes through the outer three IR blocking filters to fall on the detector housing. (The gate valve slide is highly polished and therefore emits little room-temperature radiation.) The detectors themselves do not view anything except sky beyond the cold field stop, and they are protected by two additional filters, but the increased heat flow through the housing changes the temperature offset between the detectors and the control thermometer. Second, the detector chip is mounted inside the inner housing on an alumina board that has an internal thermal time constant or diffusion time of ~84 s. This is heated somewhat by launch vibrations, and much more by vibration coupled from the gate valve drive motor. Like the detector box heat load, this temperature increase reduces the gains on all pixels by essentially equal amounts, and there is a common recovery time constant. (We now plan to replace the alumina with a gold-coated silicon board.)

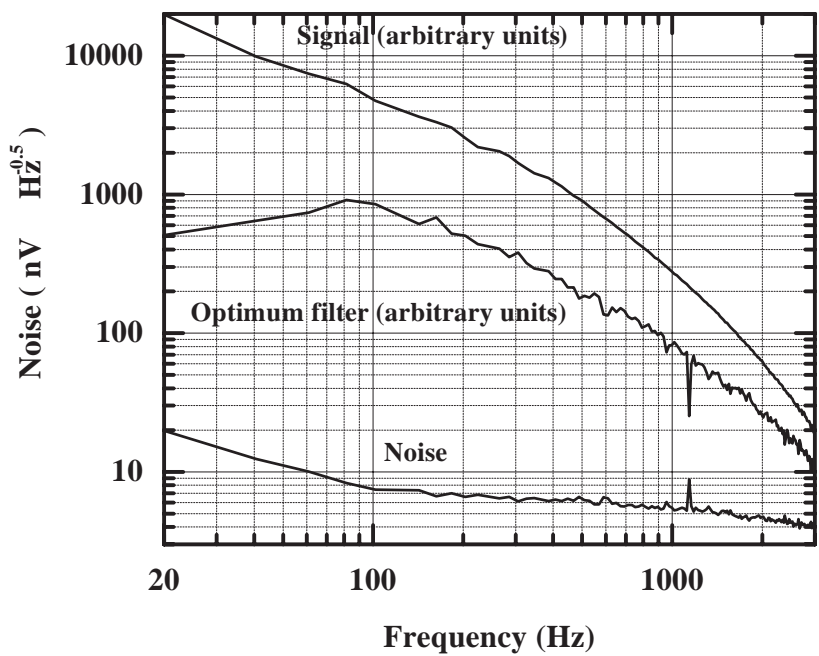

FIG. 9.—Total noise spectrum from a typical pixel, along with the X-ray pulse power spectrum and the resulting optimum filter response.

Figure 10 shows an uncorrected pulse height vs. time scatter plot for all good events observed during the flight, and covers part of the prelaunch period as well as some of the time after the experiment is on the parachute. The drop in gain during the portion of the flight with the gate valve open is apparent, as is the common recovery of most of the pixels following transient heating events. However, inspection of the gain recoveries following gate valve operations shows that a number of the pixels have longer recovery time constants than the rest, and a few are much longer. Since the pixels are all part of the same silicon chip, the different gains imply differences in temperature between the pixels themselves. There is no correlation between adjacent pixels, so it seems unlikely to be radiative heating. On the other hand, the thermal time constant of the pixels is known from the pulse recovery times to be only a few milliseconds. The mechanical vibrations of the pixels, which can be detected through the microphonic signals they induce, damp out more quickly than the pixel gains recover, so it is not energy transfer from the mechanical system. We currently think that the heat originates as mechanical input, but is stored in the small spots of silicone rubber that were placed near the base of each pixel leg to shorten the damping time for the ~1 kHz vibrations of the individual pixels, which otherwise could be minutes. These spots vary in size and exact position, and this might account for the variable recovery times.

These gain variations are corrected by making cubic spline fits to the amplitude of the potassium $K_\alpha$ events from the $^{41}$Ca continuous calibration source. The analog-filtered data are used for this despite their poorer resolution, because only every fourth calibration source event is recorded with the digital data, and the full rate is needed to track gain variations during the flight.

### 5.4 *Linearity Corrections*

The detectors have a nonlinear energy response, partly because the heat capacity increases with temperature during the pulse, but primarily because of the non-linear response function of the thermometers ($R(T) \approx R_0 \exp(T_0/T)^{0.5}$). We have used cubic polynomials in pulse height to fit data from the internal and gate-valve calibration sources taken for ~30 minutes immediately prior to the launch. Figure 11 shows the lines used in this fit. The nonlinearity at 3.3 keV is about 19%, and residuals from the cubic fit are all less than 0.5 eV below 1 keV.

### 5.5 *Livetime Determination*

The hardware deadtime during the flight was determined by counting the fraction of baseline triggers that were missed. Baseline samples are triggered by unsynchronized but stable oscillators, and thus measure the deadtime from everything except their own telemetry time. This correction was then determined by counting the number of accidental coincidence pulses that occurred on the telemetered baselines, resulting in a total hardware deadtime of 7%.

An analysis deadtime is introduced because some telemetered pulses are rejected due to pileup effects that would prevent accurate determination of the pulse height. Rejected events were counted in several energy ranges, and it was found that the analysis deadtime was 24%, independent of energy. Events that were excluded because of risetime or falltime were not included in the count because they are presumably not from X rays stopped in the active part of the detector used in determining the detector area × efficiency function.

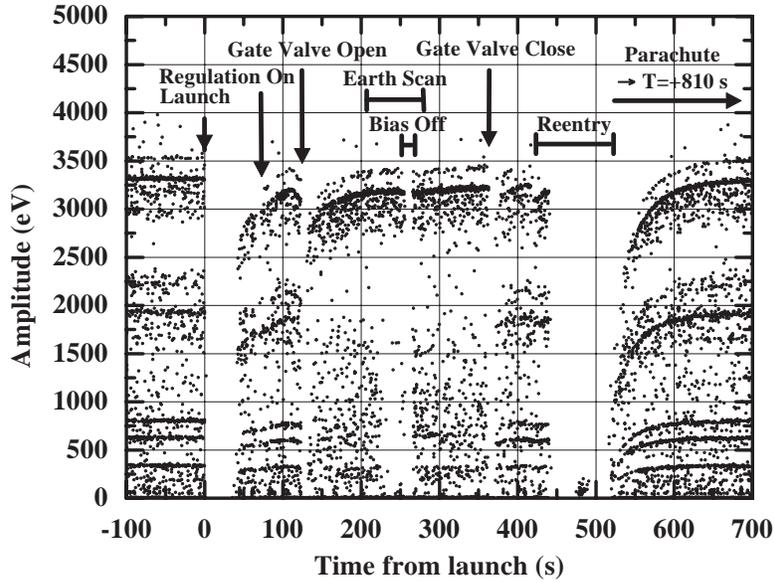

Fig. 10.—Uncorrected pulse height of all events vs. time, showing the long recovery times following gate valve operations. Lines from the gate valve calibration source can be seen when the valve is closed, and the 3.3 and 3.6 keV lines from the $^{41}$Ca source used to correct the gain variations are visible throughout. The broad band below the $^{41}$Ca lines is due to poor thermalization of X rays that penetrate the HgTe and are absorbed in the Si substrate. The HgTe is almost completely opaque below 1.2 keV, so this does not affect energies of interest. The O VII line and effects of the earth scan and the carbon-K absorption edge of the IR filters can be seen in the interval where the gate valve is open.

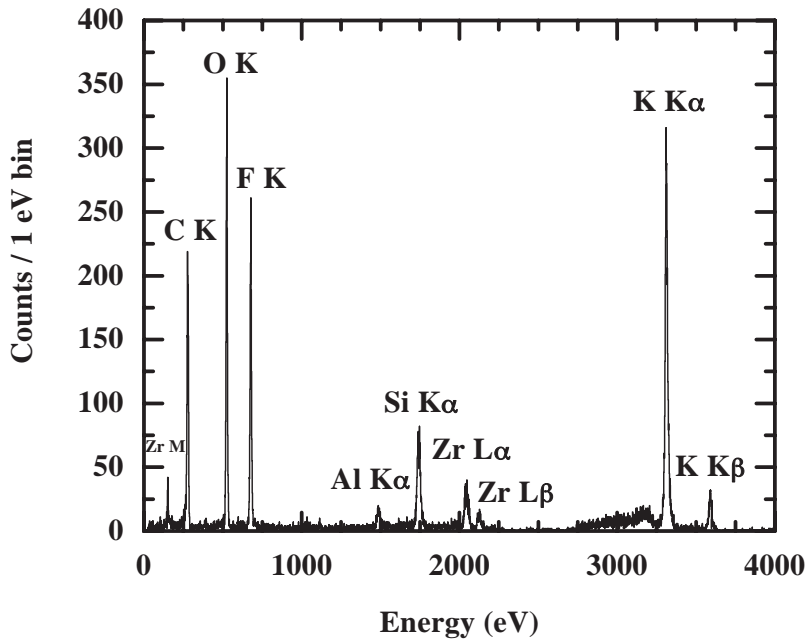

FIG. 11.—Calibration source spectrum taken during the 30 minutes immediately before launch. These lines are used to establish the energy scale, which is well fit by a cubic polynomial.

### 5.6 *Cross Checks*

An over-all sanity check is desirable, to see how well the accepted events represent the true X-ray flux. This is particularly important, since our experience with thin-window detectors on sounding rocket flights from White Sands is that one sees a significant rate from precipitating electrons on about 10% of all flights, even with strong magnetic "brooms" ahead of the collimators. The steradian field of view made a magnetic deflector impractical, so we were very concerned about possible electron contamination. This would produce an apparent continuum that would among other things distort determination of the line to continuum ratio that is one of our major science goals.

Fortunately, the expected data are strongly constrained by existing all-sky survey data. We have used maps from the UW rocket and *ROSAT* sky surveys to determine the count rate in their broad energy bands for the exact field of view of this observation. The seven bands of the rocket survey and six for the *ROSAT* survey completely overlap the energy range of the current instrument, and reasonably constrain the possible spectra. We have tried two comparison methods. In one we fit multi-component thermal and power-law spectra to the *ROSAT* and rocket data. The resulting models (see Table 3) are folded through our instrument response function and integrated over 100 eV bins to average out details of the atomic physics. The results are shown in Figure 12, with the observed spectrum integrated over the same bins, and are somewhat model-dependent. Note the large difference between the models in the lowest bin, which is due to a single iron line complex at 70 eV. The observed rates are generally consistent within the model uncertainties over the range of interest.

The good energy resolution allows a model-independent comparison to be made simply by dividing the observed pulse-height spectrum by the instrument throughput at the corresponding energy to get an approximation of the input spectrum. This spectrum can be folded through the sky survey band responses. This does not work well in the lowest energy bands where the sky survey response changes appreciably within the detector resolution. For the *ROSAT* R4, R5, R6, and R7 bands, however, it gives respectively 96%, 82%, 70% and 88% of the observed rates from the sky survey. This is consistent with the implications of Figure 12, given that the dip in the data there corresponds to the middle of the R6 band.

The entire field of view was occulted by the dark earth for 20 s near the middle of the flight. During this time, 6 events were observed in the 25 - 1000 eV range while 6.3 were expected from internal background due to the calibration source and ~0.06 from cosmic rays. Together, these results give us some confidence that we understand the entire system response and that there is no evidence for significant electron contamination.

TABLE 3
Model Parameters for Fits to Soft X-ray Diffuse Background Sky Survey Data

| Parameter | Thermal Components with Solar Elemental Abundances[a] | Thermal Components with Depleted Elemental Abundances[b] |
|---|---|---|
| Power Law Component: | | |
|   Exponent | 1.52 | 1.52 |
|   Normalization | 12.3 | 12.3 |
|   $N_H$ (H atoms cm$^{-2}$) | $1.8 \times 10^{20}$ | $1.8 \times 10^{20}$ |
| Absorbed Thermal Component: | | |
|   Temperature (kT in keV) | 0.225 | 0.317 |
|   Emission Measure (cm$^{-6}$ pc) | 0.0037 | 0.0053 |
|   $N_H$ (H atoms cm$^{-2}$) | $1.8 \times 10^{20}$ | $1.8 \times 10^{20}$ |
| Unabsorbed Thermal Component: | | |
|   Temperature (kT in keV) | 0.099 | 0.107 |
|   Emission Measure (cm$^{-6}$ pc) | 0.0088 | 0.053 |

[a] Anders & Grevesse (1989)

[b] from Savage & Sembach (1996) for cool clouds towards ζ Ophiuchi (their Table 5)

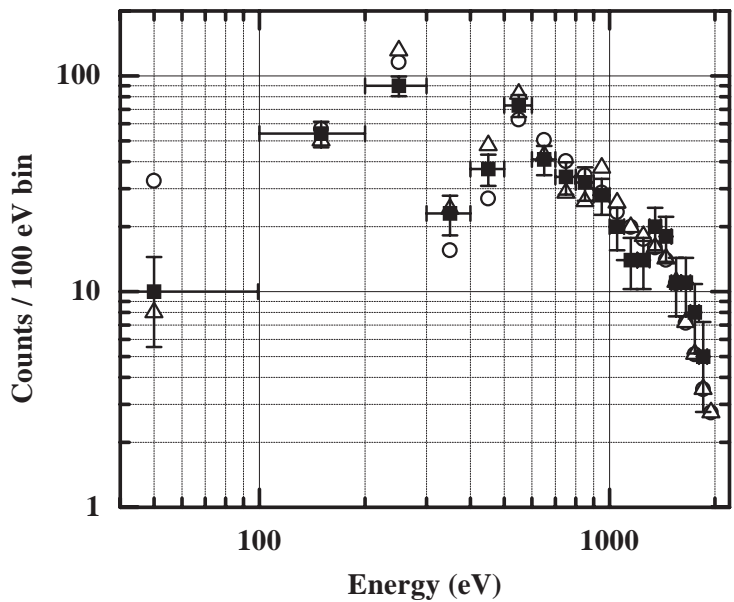

Fig. 12.—Check of absolute rates. The flight pulse height data are integrated over 100 eV bins and compared to predictions from models fit to *ROSAT* and Wisconsin sky survey data from the target area. Model parameters are given in Table 3. Circles show the solar abundance model, triangles show depleted abundances.

The geocoronal He II Ly-$\alpha$ line at 41 eV is an additional potential contaminant. It is produced by fluorescence from direct and scattered solar radiation, so its intensity varies strongly across the night sky, but is generally in the 1–10 Rayleigh range. The organic substrates in our IR blocking filters have a calculated transmission of $\sim 10^{-8}$, however, so less than one count is expected for 10 R. We observe only two events in the 25 - 55 eV energy range, while 1.3 are expected from internal background.

## 6. DISCUSSION
### 6.1 *Line Identifications and Fluxes*

Figure 13 shows the net spectrum from ~100 s of livetime on the target region, along with the solar abundance, two-temperature equilibrium thermal emission plus power law model spectrum given in Table 3. The Lyman alpha lines of C VI and O VIII are clearly present, as are the helium-like triplet of O VII and the Fe-M line complex at 70 eV. Derived intensities for these lines corrected for instrument response are given in Table 4. The $4.8 \pm 0.8$ photons cm$^{-2}$ s$^{-1}$ sr$^{-1}$ total for the O VII triplet intensity can be compared to previous values deduced from lower-resolution observations of similar parts of the sky: Inoue et al. (1979) obtained $8.3 \pm 0.6$ photons cm$^{-2}$ s$^{-1}$ sr$^{-1}$ with a single-line fit to all of the flux observed from 0.5–1 keV with a gas scintillation proportional counter, Rocchia et al. (1984) show a fit to a spectrum obtained with a solid state detector that includes an O VII line with an intensity of about 4 photons cm$^{-2}$ s$^{-1}$ sr$^{-1}$, Gendreau et al. (1995) found $2.3 \pm 0.3$ photons cm$^{-2}$ s$^{-1}$ sr$^{-1}$ for a composite spectrum taken with the ASCA SIS CCD cameras, and Mendenhall & Burrows (2001) obtained $4.2 \pm 1.8$ photons cm$^{-2}$ s$^{-1}$ sr$^{-1}$ with a sounding rocket CCD observation. Gendreau et al. (1995) also fit $0.6 \pm 0.15$ photons cm$^{-2}$ s$^{-1}$ sr$^{-1}$ for the O VIII line.

The fortuitously located aluminum L-edge in the IR blocking filter transmission provides a small but significant sensitivity (~0.3%) to the very bright Fe IX,X,XI line complex at 70 eV. There is a clear feature with four events at this energy (P < 0.0004), where 0.3 counts are expected from continuum and internal background, and 28 counts from the iron lines in the solar-abundance model. The asymmetric wing on the low side of the feature in the model is produced by lines of Fe XII near 64 eV, consistent with the fifth count observed at this energy. The line complex is produced by four successive ionization states, so the total intensity is not particularly sensitive to details of the model. The observed faintness is therefore difficult to explain except as an underabundance of iron in the gas phase, presumably due to the survival of grains in the hot gas responsible for this part of the diffuse background. Previous measurements have placed upper limits of 0.15–0.3 on the depletion factor (Bloch, Priedhorsky, & Smith 1990; Jelinsky, Vallerga, & Edelstein 1995, Vallerga & Slavin 1998). We nominally find 0.14 of the flux predicted for solar abundances, with large statistical and systematic uncertainties. The lack of observed iron lines near 800 eV is also consistent with this depletion, but in the model they are produced by a different thermal component and it may be possible to explain their weakness by adjustments in the model parameters.

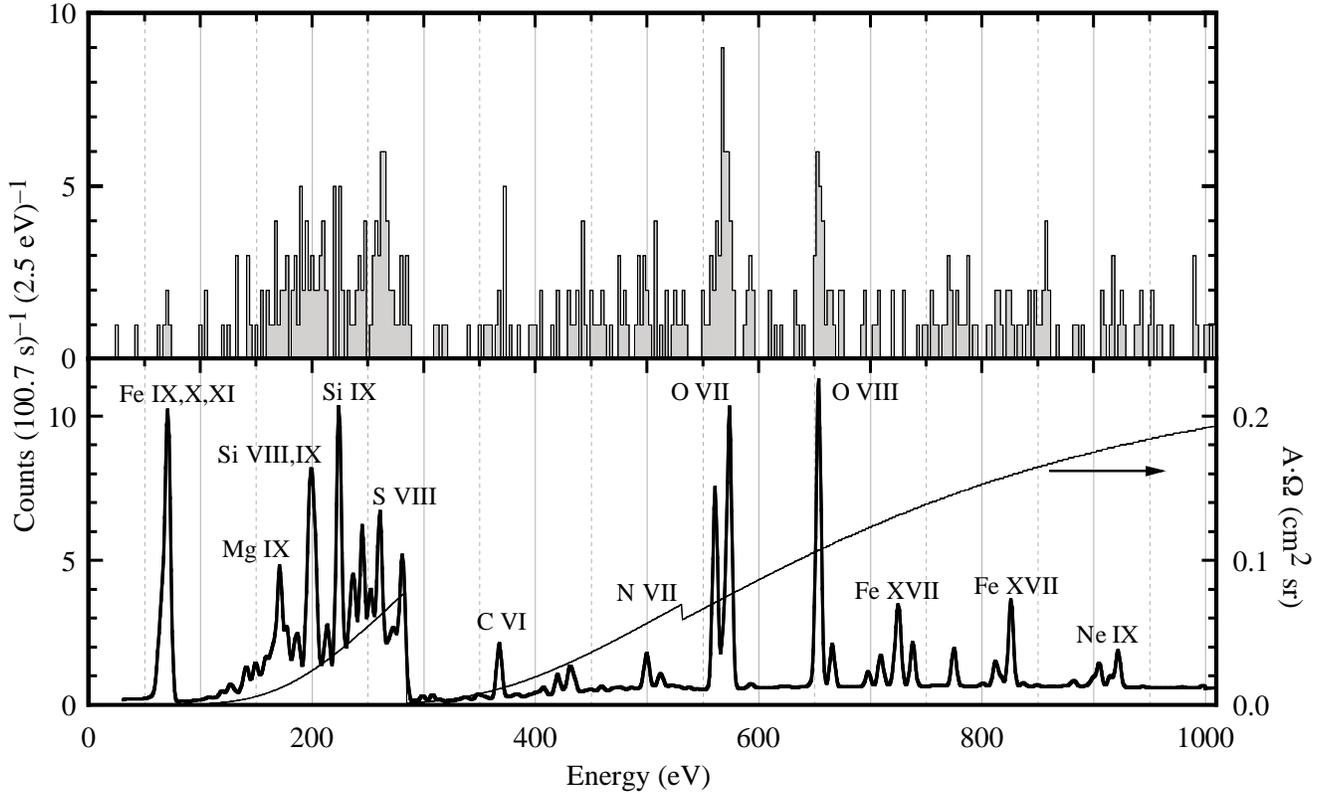

Fig. 13.—Observed pulse height spectrum from target region (top) and the normal abundance two temperature thermal plus power law model given in Table 3 (bottom). The shape of the throughput (A•Ω) curve in the lower panel is determined primarily by the transmission of the infrared blocking filters. The model as shown includes this instrument response.

TABLE 4
Observed Line Fluxes

| Line Identification | Line Energy (eV) | Photon Flux (photons cm$^{-2}$ s$^{-1}$ sr$^{-1}$) | Energy Flux (ergs cm$^{-2}$ s$^{-1}$ sr$^{-1}$) |
| --- | --- | --- | --- |
| Fe IX,X,XI | 69–72 | 100 ± 50 | 1.1 ± 0.6 × 10$^{-8}$ |
| C VI | 368 | 5.4 ± 2.3 | 3.2 ± 1.2 × 10$^{-9}$ |
| O VII (triplet) | 561–574 | 4.8 ± 0.8 | 4.4 ± 0.7 × 10$^{-9}$ |
| O VIII | 653 | 1.6 ± 0.4 | 1.7 ± 0.4 × 10$^{-9}$ |

On the other hand, the very brief observation of the North Polar Spur at the end of the 360° Earth scan shown in Figure 14 shows evidence for a bright Fe XVII line (P <0.02 for 4 counts in a known position). The three-count peak at 750 eV also would be significant if a line were expected at this energy, but there is ~30% probability of a random feature of this size somewhere in the 50 to 1000 eV range. The larger peak at 990 eV is just significant at the 5% level, and there is a small peak at the corresponding location in the target area spectrum, but serious speculation should wait for additional data.

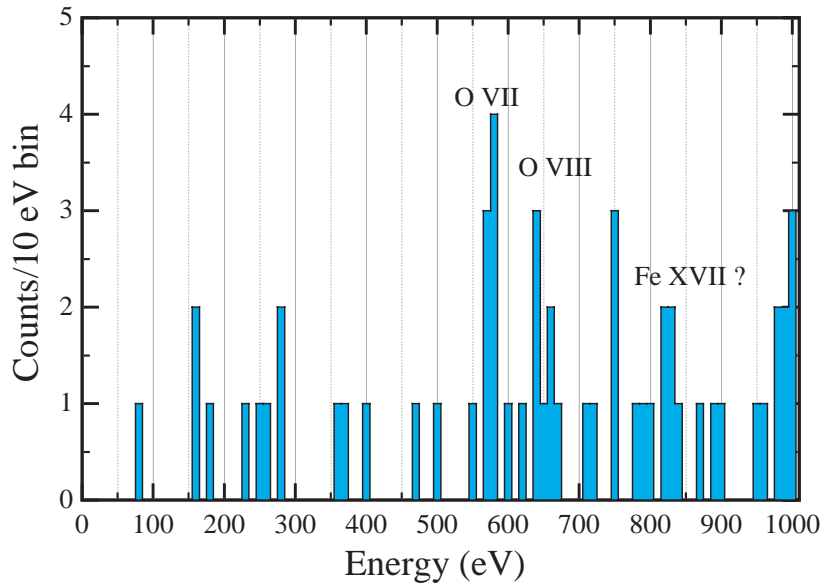

Fig. 14.—Data from a 12 s exposure at the end of the earth scan. Most of this overlaps the bright North Polar Spur 3/4-keV enhancement.

### 6.2 Components of the 3/4 keV Background

We can easily calculate the contribution of the oxygen lines observed on this flight to the band rates of the *ROSAT* diffuse background survey. The total contributions from our measured oxygen line fluxes are shown as solid red bars descending from the *ROSAT* survey points in Figure 15. The largest is to the *ROSAT* R4 band, plotted at 0.65 keV, where it accounts for 32% of the total observed rate. In an attempt to determine the absolute minimum amount of additional thermal radiation that would have to come from the plasma producing these lines, we have calculated emission models where all the other metals are fully depleted, and the temperature is optimized to produce the largest possible fraction of the total flux in the oxygen lines. The contribution of the rest of the radiation from this model, which is primarily recombination continuum, was calculated for each survey band and is shown as the pink extension to the line contribution. In the R4 band it amounts to an additional 10%. We can therefore say that at least 42% of the observed diffuse background in the R4 band comes from thermal emission located at $z < 0.01$, due to the observed positions of the oxygen lines at their laboratory wavelengths. It can also be shown that 38% of the R4 band is accounted for by the AGN directly observed by deep surveys with *ROSAT* (Hasinger et al. 1993; Hasinger 2000, private communication) and *Chandra* (Mushotzky et al. 2000), nominally leaving 20% for any truly diffuse extragalactic contribution.

This is significant because Cen & Ostriker (1999) have predicted that a large fraction of the truly diffuse background flux observed in this spectral region should come from redshifted lines emitted by hot intergalactic structures containing most of the "missing baryons." Unfortunately, the very small number of counts in the current observation results in fairly large statistical errors on the brightness of individual lines. Using a 2σ lower limit to the observed oxygen line fluxes, we can allow as much as 34% for extragalactic diffuse emission. Modest amounts of additional data will quickly reduce the statistical uncertainties in this number. The AGN contribution is unlikely to increase by as much as 4%, since the faintest AGN observed in the Chandra deep survey are so heavily absorbed that they contribute very little to the R4 band (Mushotzky, private communication). However, the lower limit to the thermal contribution could increase substantially as better statistics allow the identification of unshifted lines of other elements. Much longer observations could make a direct search for the redshifted oxygen emission lines from individual intergalactic filaments.

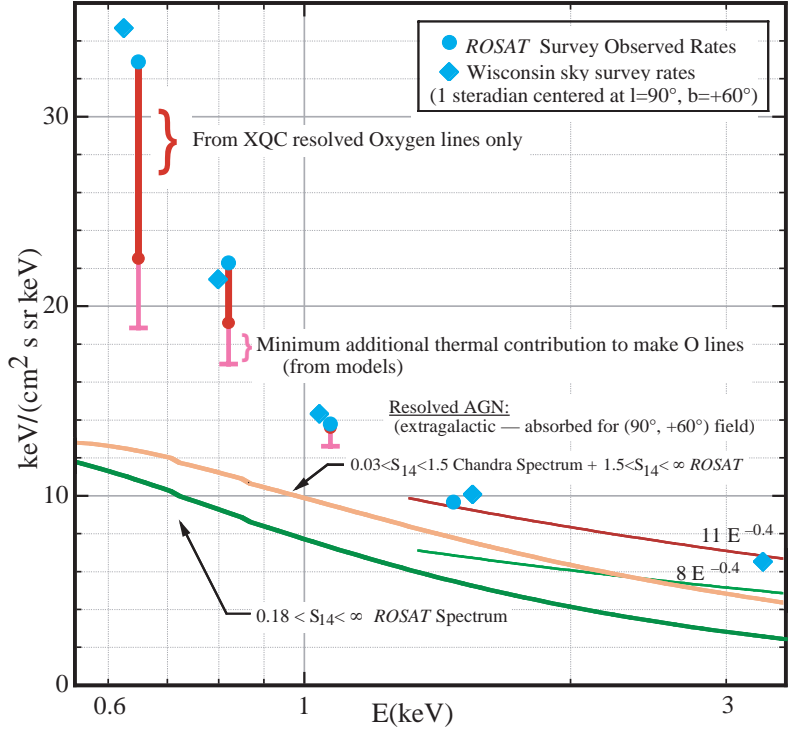

FIG. 15.—Contributions to the observed total diffuse background. The blue circles and diamonds show the total observed sky survey rates. The red bars extending down from these are the contribution to these band rates from the O VII and O VIII fluxes observed in this experiment. The pink extensions on these bars are the almost model-independent minimum additional thermal emission that must be associated with the plasma producing the O lines. The upper AGN spectrum (orange line) is the total flux that has currently been resolved into discrete sources. The gap between this and the bottom of the bars represents an absolute upper limit to any truly diffuse extragalactic source. The lines labeled 8 and 11 $E^{-0.4}$ represent lower and upper limits to the total extragalactic background in the 2-10 keV range, as determined by several experiments (Mendenhall & Burrows 2000, and references therein).

The origin of this 40-60% of the R4 band background that is nearby thermal emission is not currently known. It does not appear that the low-density cavity around the Sun contributes appreciably at these energies (Snowden et al. 1993). It could be due to a hot halo, but if this is the only significant emitter, it needs to have a rather small scale height. The observed total intensity is the same in the Galactic plane, where the ≥ 40% extragalactic contribution is completely absorbed by interstellar gas, so the local contribution needs to be about twice as large at low latitudes as at high. This requires either a scale height much less than the mean absorption length in the disk (about 700 pc), or yet another independent source within the disk. In either case the spectrum must show major variations from high latitudes, where it is ~40% extragalactic continuum, to the plane, where it should be essentially all in lines.

### 6.3 *The 1/4 keV Spectrum*

The spectrum below the carbon edge at 284 eV is very crowded, and at our resolution shows only blends of lines. This wavelength region was observed at similar resolution with the Bragg crystal Diffuse X-ray Spectrometer (DXS) experiment, where exhaustive efforts to fit the spectrum with existing emission models, including non-equilibrium effects, multiple temperatures and variable abundances, were notably unsuccessful (Sanders et al. 2001). Real spectral analysis in this region, which contains most of the emission from $10^{6.0}$ K gas, will require resolutions of 2 eV or better that allow analysis of individual line intensities and ratios where the atomic physics is well understood.

Figure 16 shows this part of the spectrum from our current observation overlaid with the DXS spectrum from a different region of the sky. The overall intensity is normalized using *ROSAT* observations of the two areas, and has been corrected to the sounding rocket instrument response. The spectra are different, but share some of the major features that made the DXS spectrum difficult to fit. In both cases, agreement is improved by using standard cold-gas depletions, which reduce the strong Si and Mg lines and enhance the S VIII feature, but significant problems remain.

The DXS observation was at low Galactic latitudes, where the X rays presumably all come from the local low-density cavity around the Sun since the mean free path at these energies is only about $1 \times 10^{20}$ H cm$^{-2}$. The region observed on this flight includes some of the high latitude enhancements that are three to four times brighter than the in-plane background (see Figure 6). The location of the additional emission is uncertain: broad-band spectra show no evidence for hardening due to absorption by the disk gas, arguing that the enhancements are part of the local cavity (Snowden et al. 1990), but shadowing experiments show that in some directions half of the high-latitude 1/4–keV background originates at $z > 300$ pc (Burrows & Mendenhall 1991; Snowden et al. 1994). Progress on understanding this problem has been slow (Kuntz & Snowden 2000), but high spectral resolution observations will provide an important additional key when they are available from more parts of the sky.

### 6.4 *An Accidental Experiment*

An additional serendipitous result comes from the thin filters, the low energy threshold of the detectors, and their inherent sensitivity to non-ionizing events. One solution to the small-scale structure problem in cold dark matter models is to postulate that the dark matter is strongly interacting (Spergel & Steinhardt 2000). An attractive candidate for this type of particle can apparently be ruled out for a wide range of masses extending down to 0.5 a.m.u. because it would have produced more than the observed number of counts in our detectors between 25 and 60 eV due to coherent scattering from detector nuclei (Wandelt et al. 2001).

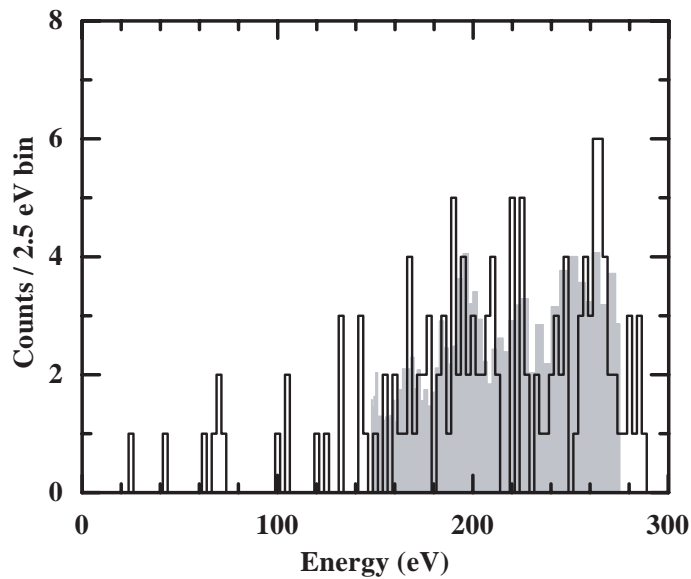

FIG. 16.—Low-energy portion of pulse height spectrum from this experiment compared with a Bragg crystal spectrometer spectrum from the Diffuse X-ray Spectrometer observation of a region near the Galactic plane (shaded; Sanders et al. 2001). The DXS spectrum has been corrected to the sounding rocket instrument response and the overall amplitude normalized using the ratio of rates in the respective regions from the *ROSAT* sky survey R2 band. The average spectral resolutions of these instruments are similar, with the crystal spectrometer better at 150 eV and worse at 284 eV. Statistical errors for the DXS data are much smaller than those for the rocket.

# 7. CONCLUSIONS

Despite the scant two minutes of observing time and 1/3 cm$^2$ collecting area on this flight, the high spectral resolution has produced some interesting results: iron-containing grains seem to have survived in the hot gas, the spectrum of the North Polar Spur is different in this respect, and, as an extreme upper limit, < 34% of the high-latitude X-ray background at these energies can have a diffuse cosmological source. The low threshold energy and lack of shielding have also made it possible to rule out strongly interacting dark matter candidates with masses larger than ~ 500 MeV. Similar observations promise much more of interest with increases in observing time and spatial coverage. For the next sounding rocket flight, we should be able to increase the area of each pixel by a factor of four without appreciably increasing the heat capacity, and observing time on the sky can be more than doubled. We also have found that increasing the thermistor thickness will eliminate most of its excess noise (Han et al. 1998), which should improve the energy resolution by about a factor of two (McCammon et al. 2002). Beyond that, improvements in thermometer technology now underway in various laboratories promise resolutions better than 2 eV.

It is impractical to list all of the graduate and undergraduate students who have participated in the development of the detectors, refrigerator, and other hardware used on this flight, but we are grateful to all of them for their efforts. We would like to recognize the major contributions of Brad Edwards, Sang-In Han, Dana Peters, Paul Plucinsky, Mark Devlin, Tam Helfer, Yvonne Marietta, Steve Nahn, Yoji Natori, Denise Steffensrud, and Britt Thesen. We thank Bob Baker, Kevin Boyce, Jim Caldwell, Norm Dobson, Jeff DuMonthier, Phil Goodwin, Chuck Hanchak, Ken Simms, George Winkert, Dale Arbogast, John Gygax, Scott Murphy, Frank Shaffer, Regis Brekosky, Richard McClanahan, Brent Mott, Carol Sappington, Carl Stahle, and Malcolm Bevan for their help with various aspects of the detectors and payload. We greatly appreciate the support of the sounding rocket staff at Wallops Flight Facility and White Sands Missile Range. We thank Dave Burrows for his careful reading of the manuscript. This work was supported in part by NASA grant NAG5-679.